\documentclass[aps, prb, reprint, groupedaddress, amsmath, amssymb]{revtex4-1}

\usepackage[utf8]{inputenx}
\usepackage[T1]{fontenc}
\usepackage{lmodern}

\usepackage{xcolor, graphicx}
	\graphicspath{{../figures/}}
\usepackage{siunitx}
\usepackage[version=4]{mhchem}
\usepackage{multirow}
\usepackage[pdfborder={0 0 0}]{hyperref}
	\hypersetup{
		pdfauthor = {Angelo Peronio and Franz J. Giessibl},
		pdftitle  = {Attempts to test an alternative electrodynamic theory of superconductors by LT STM and AFM}}
	
\newcommand{\vm}{V_\mathrm{m}}
\newcommand{\fm}{f_\mathrm{m}}
\newcommand{\de}{\mathrm{d}}
\newcommand{\vv}[1]{\mathbf{#1}}
\DeclareMathOperator{\rot}{rot}
\DeclareMathOperator{\grad}{grad}
\newcommand{\lambdaL}{\lambda_\mathrm{L}}
\newcommand{\lambdaTF}{\lambda_\mathrm{TF}}
\newcommand{\lambdaEff}{\lambda_\mathrm{eff}}
\newcommand{\kts}{k_\mathrm{ts}}
\newcommand{\bes}{\mathcal{I}_0}

\begin{document}

\title{Attempts to test an alternative electrodynamic theory of superconductors\texorpdfstring{\\}{} by low-temperature scanning tunneling and atomic force microscopy}

\author{Angelo Peronio}
\email{angelo.peronio@ur.de}
\affiliation{Institut für Experimentelle und Angewandte Physik, Universität Regensburg, D-93053 Regensburg, Germany}

\author{Franz J. Giessibl}
\affiliation{Institut für Experimentelle und Angewandte Physik, Universität Regensburg, D-93053 Regensburg, Germany}

\date{August 31, 2016}

\begin{abstract}
We perform an experiment to test between two theories of the electrodynamics of superconductors: the standard London theory and an alternative proposed by J.~E. Hirsch [\href{http://dx.doi.org/10.1103/PhysRevB.69.214515}{Phys. Rev. B \textbf{69}, 214515 (2004)}]. The two alternatives give different predictions with respect to the screening of an electric field by a superconductor, and we try to detect this effect using atomic force microscopy on a niobium sample. We also perform the reverse experiment, where we demonstrate a superconductive tip mounted on a qPlus force sensor. Due to limited accuracy, we are able neither to prove nor to disprove Hirsch’s hypothesis. Within our accuracy of 0.17 N/m, the superconductive transition does not alter the atomic-scale interaction between tip and sample. 
\end{abstract}

\maketitle
\section{Introduction}
The first phenomenological description of superconductivity was provided by the brothers Fritz and Heinz London in 1935,\cite{London1935, London1935a, London1950} in which they postulated that part of the electrons in a superconducting body obey two simple equations. The first one,
\begin{equation} \label{eq:london1}
\partial_t \, \vv{j_s} = \frac{n_s e^2}{m} \, \vv{E},
\end{equation}
expresses the free, collisionless acceleration of the superconducting charge carriers under the action of an electric field. Here $n_s$, $\vv{j_s}$, and $m$ are the number density, current density and mass of the superconducting electrons, and SI units are used. The second equation,
\begin{equation} \label{eq:london2}
\rot \vv{j_s} = - \frac{n_s e^2}{m} \, \vv{B}
\end{equation}
leads to the Meissner effect: the expulsion of the magnetic field from the interior of a superconductor. A proper microscopic theory of superconductivity appeared only in 1957 with Bardeen, Cooper, and Schrieffer,\cite{Bardeen1957} and within this framework the London equations describe the limit where the response to electric and magnetic fields is local.

As discussed by J. E. Hirsch,\cite{Hirsch2004} the London equations present two difficulties. First, they predict that an accumulated space charge should persist for arbitrarily long times as the temperature approaches absolute zero or the critical temperature $T_c$,\cite{*[] [{, p. 60.}] Rickayzen1965} a phenomenon that, to our knowledge, has never been observed experimentally. Second, they predict that an electromagnetic wave is exponentially damped inside a superconductor with a characteristic length $\lambdaL$, the London penetration depth. This description cannot be valid in the low-frequency limit, since a static electric field inside a superconductor will generate an infinite current, as per Eq.~\eqref{eq:london1}.

To solve these difficulties Hirsch follows an early attempt of the London brothers,\cite{London1935, London1935a} and replaces Eq.~\eqref{eq:london1} with
\begin{equation} \label{eq:hirsch}
\partial_t \, \vv{j_s} = \frac{n_s e^2}{m} \, (\vv{E} + \grad \phi),
\end{equation}
where $\phi$ is the electric potential.\cite{Hirsch2004} In this formulation a static electric field can exist inside a superconductor without generating any electrical current.

\begin{figure}
	\includegraphics{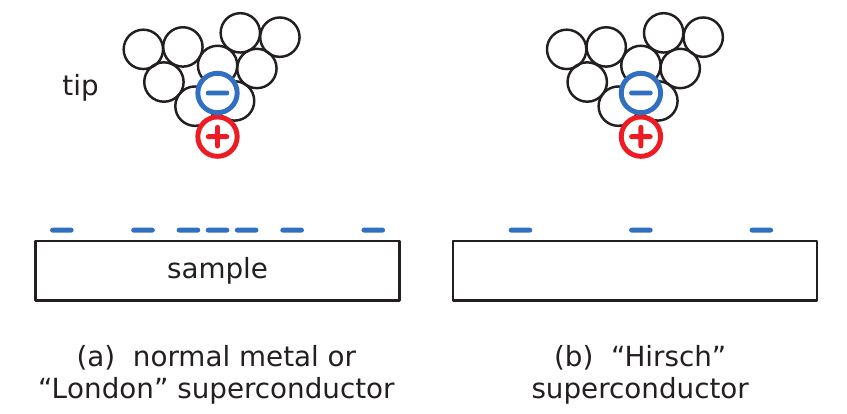}%
	\caption{\label{fig:dipole} The response of a superconductor to the electric field of an AFM tip apex. A ``London'' superconductor screens an applied field like a normal metal, within the Thomas-Fermi screening length of about \SI{0.1}{nm}. The screening of an ``Hirsch'' superconductor is instead much weaker, with a characteristic length of \SI{39}{nm} for niobium at zero temperature.}
	
\end{figure}

To decide between these two theories, one can consider what happens when an electrostatic field is applied to a superconductor, as first proposed in Ref.~\onlinecite{Hirsch2015}. Figure~\ref{fig:dipole} depicts the situation, in which an atomically sharp metal tip is approached to a superconductive sample. %
The electron cloud of the tip does not follow the sharp curvature of the tip apex, instead it smooths out -- the so-called Smoluchowski effect\cite{Smoluchowski1941, Teobaldi2011, Ellner2016} -- giving rise to an electric dipole located at the apex. The sample responds to the dipole's field by piling up surface charge, in order to have no electric field in its interior. A ``London'' superconductor behaves in this respect like a normal metal, where the spatial extent of this screening is the Thomas-Fermi screening length $\lambdaTF$ -- about \SI{0.1}{nm}.\cite{Kolacek2002} In other words, the charge density which accumulates on the surface cannot change over distances smaller than $\lambdaTF$. From Hirsch's equation~\eqref{eq:hirsch} it can instead be shown that this characteristic length should be the much larger London penetration depth\cite{Hirsch2004} -- \SI{39}{nm} for Nb.\cite{Maxfield1965} The spatial extent of the accumulated charge density is on the order of the tip-sample distance,\cite{Hirsch2015} so if the latter is smaller than $\lambdaL$, an ``Hirsch'' superconductor will not be able to pile up surface charge as tightly as a normal metal, and the electrostatic force between tip and sample will be different in the ``Hirsch'' and ``London'' cases.

In our experiment we combined atomic force microscopy (AFM) and scanning tunneling microscopy (STM) to measure the interaction between a metal tip and a niobium surface. In particular, we looked for differences between measurements performed below and above the critical temperature $T_c = \SI{9.25}{K}$ of the sample,\cite{crc} which could be due to the physics predicted by Hirsch.

\section{Methods}
We employed a combined STM/AFM (Omicron LT STM/AFM, Omicron Nanotechnology) cooled by an helium bath cryostat to \SI{4.4}{K} and operated in ultrahigh vacuum (UHV) at a base pressure of \SI{3e-9}{Pa}. The Nb(110) sample (MaTecK GmbH, purity 99.99\%) was prepared by repeated cycles of \ce{Ar+} sputtering and annealing up to \SI{1170}{K}, resulting in a reconstructed surface due to oxygen segregating from the bulk.\cite{Suergers2001} A subsequent brief sputtering removed this reconstruction, leaving a surface with \si{nm}-scale asperities. Additional measurements involved a Cu(111) and a Cu(110) sample (MaTecK GmbH, purity 99.9999\%), prepared by repeated sputtering and annealing up to \SI{785}{K}. We used an etched tungsten tip, prepared by field evaporation in UHV and \emph{in situ} poking into a clean copper sample, likely resulting in a copper-coated tip apex.\cite{*[] [{, supplemental material.}] Hofmann2014} The tip is mounted on a qPlus sensor\cite{Giessibl1998} operated in frequency-modulation mode,\cite{Albrecht1991} with a quality factor at low temperature ranging from \SI{250000}{} to \SI{540000}{}.\footnote{In our system the quality factor of the sensor depends on the coupling between the sensor holder and its receptacle on the scanner, and changes if the former is slightly displaced.} The tip-sample interaction is detected via the frequency shift $\Delta f$ of the sensor from its unperturbed resonance frequency $f_0 = \SI{47388}{Hz}$, which is related to the gradient of the vertical force between tip and sample. Precisely, $\Delta f = \frac{f_0}{2 k} \, \langle \kts \rangle$, where $k = \SI{1800}{N/m}$ is the stiffness of the sensor, $\kts = - \partial_z F_z$ is the local ``spring constant'' of the tip-sample force, and the angle brackets indicate a weighted average over the oscillation amplitude $A$ of the tip\cite{Welker2012, Giessibl1997}$^,$\footnote{The $\Delta f(z)$ spectra could be deconvolved to get the tip-sample force and force gradient,\cite{Sader2004, Welker2012} but doing so would increase the noise.}
\begin{equation}
\langle \kts \rangle = \frac{2}{\pi A^2} \int_{-A}^{A} \de z \; \kts(z) \,  \sqrt{A^2 - z^2}.
\end{equation}
For the STM measurements a bias voltage $V$ was applied to the sample, and the tunneling current $I$ was measured by a DLPCA-200 transimpedance amplifier (FEMTO Messtechnik GmbH) connected to the tip.

\section{Results and discussion}
\subsection{Magnitude of the effect}
How big is the signal that we expect to measure? For the normal metal of Fig.~\ref{fig:dipole}(a) the electrostatic part of the tip-sample interaction can be thought as the attraction between two dipoles: the dipole of the tip and its image dipole in the sample. After the superconductive transition this interaction will be still present in the ``London'' case, and will instead be strongly reduced in the ``Hirsch'' superconductor of Fig.~\ref{fig:dipole}(b). If Hirsch is right, the measurements above and below the critical temperature will differ, at most by the force between two aligned dipoles\cite{*[] [{, problem 4.8.}] Griffiths1998}
\begin{equation}
F_z = -\frac{12}{4 \pi \epsilon_0} \, \frac{p^2}{(2z)^4} .
\end{equation}
Here $z$ is the tip-sample distance, and $p$ is the dipole of the tip, estimated to be \SI{0.5}{D}\cite{Gross2014} or \SI{0.9}{D}\cite{Schneiderbauer2014} for copper tips.\footnote{1 debye (D) is approximatively \SI{3.34e-30}{\coulomb\metre}} The corresponding frequency shift 
\begin{equation} \label{eq:df}
\Delta f = -\frac{f_0}{2 k} \, \left\langle  \partial_z F_z \right\rangle  = -\frac{f_0}{2 k} \; \frac{3}{4 \pi \epsilon_0} \, p^2 \left\langle z^{-5} \right\rangle 
\end{equation}
is depicted in Fig.~\ref{fig:df}(a).

This is an upper estimate, accurate for $z \ll \lambdaL$ and $T \ll T_c$. If instead $T \lesssim T_c$, both the normal and the superconducting electrons will contribute to the screening -- the superconducting electrons with a characteristic length $\lambdaL$, and the normal ones with $\lambdaTF$. The result is an effective screening length $\lambdaEff$, defined in Eq.~(33) and Fig.~3 of Ref.~\onlinecite{Hirsch2012}. For niobium $\lambdaEff(\SI{4.4}{K}) = \SI{0.44}{nm}$ and $\lambdaEff(\SI{2.4}{K}) = \SI{1.48}{nm}$, significantly smaller than $\lambdaEff(\SI{0}{K}) = \lambdaL = \SI{39}{nm}$. Thus at the temperatures we are able to reach we expect a much smaller effect than what Eq.~\eqref{eq:df} predicts, but it is not easy to give a quantitative lower estimate.

\subsection{Measurements at \SI{4.4}{K}}
\begin{figure}
\includegraphics{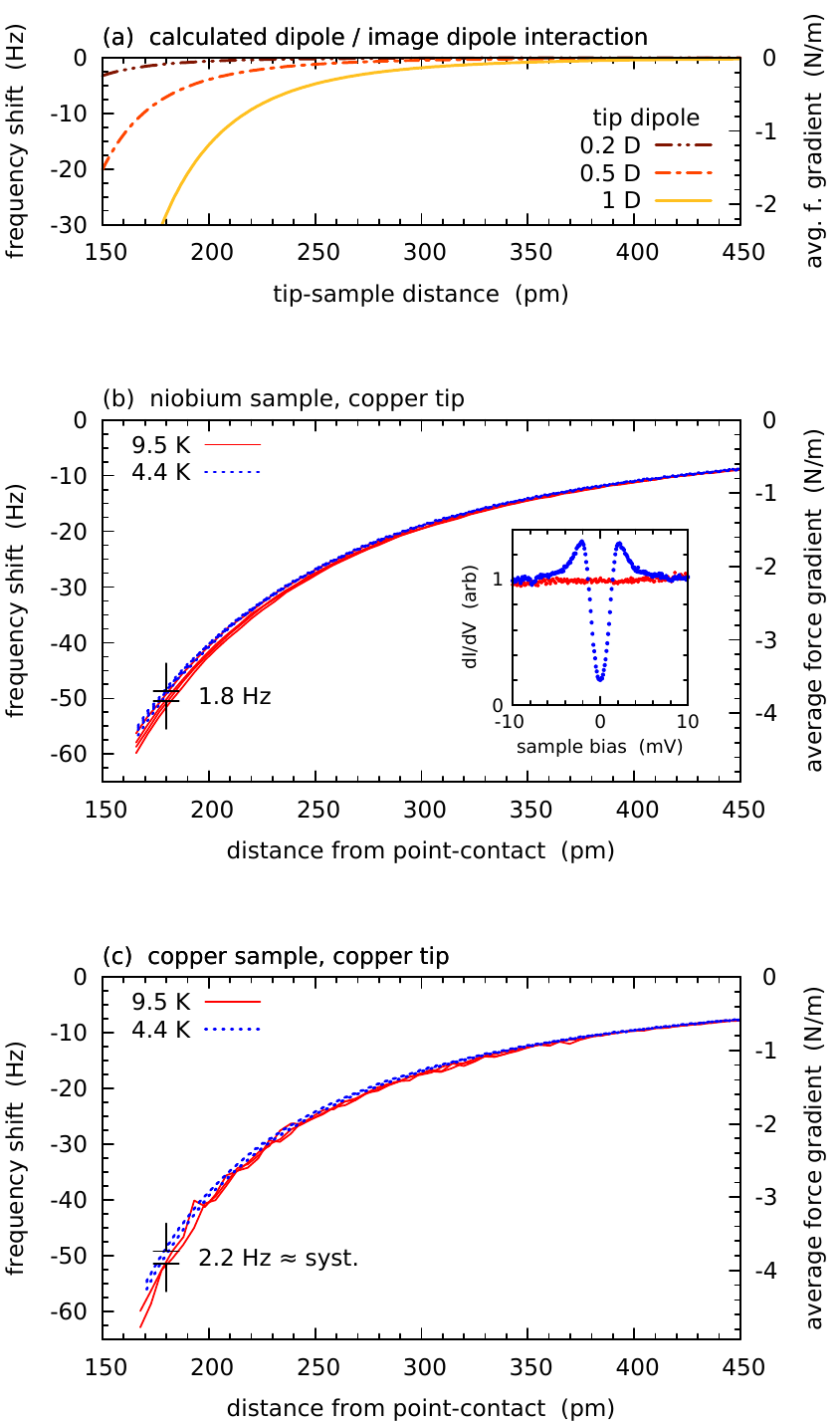}%
\caption{\label{fig:df} (a) Frequency shift due to the dipole / image dipole interaction calculated for different tip dipoles, and for the oscillation amplitude $A = \SI{50}{pm_{pk}}$ used in the experiments of panels (b) and (c). (b) $\Delta f(z)$ spectra at two different temperatures on Nb(110) and (c) on Cu(110). Only at $T = \SI{4.4}{K}$ the Nb sample superconducts, as shown by the $\de I / \de V$ spectroscopy of the superconductive gap (inset). The $\Delta f(z)$ spectra at the two temperatures are different on Nb, but this effect cannot be attributed to ``Hirsch'' superconductivity, since it is observed also on Cu. These spectra are acquired on the same point on the surface, and multiple measurements are shown. The measurements of panel (b) have been acquired in different heating-cooling cycles. The $\de I / \de V$ spectra are acquired at a tunneling setpoint $V = \SI{-20}{mV}$, $I = \SI{200}{pA}$ with a modulation voltage $\vm = \SI{300}{\micro V_{pk}}$ at $\fm = \SI{407}{Hz}$.}
\end{figure}

Figure~\ref{fig:df}(b) compares frequency-shift curves as a function of the vertical position $z$ of the tip above the niobium surface. The curves are acquired over the same atomic-scale feature at $T = \SI{4.4}{K}$ and $T = \SI{9.5}{K}$, and $\de I / \de V$ spectroscopy of the superconductive gap (inset) shows that the sample superconducts only at $T = \SI{4.4}{K}$. In order to compare these measurements to the theoretical estimate of panel (a), we need to set the zero of the $z$-axis, i.~e. we need to estimate the position of the surface. To this end, we employed a commonly-used approximation, assuming $z = 0$ at the ``point-contact'', where the tunneling conductance would reach $G_0 = 2e^2/h \approx \SI{77.5}{\micro S}$ with a non-oscillating sensor.\cite{Gimzewski1987, Schneiderbauer2014}

The measurements acquired at the two temperatures differ slightly but reproducibly -- different traces correspond to different repetitions -- and indeed below the transition temperature the tip-sample interaction is weaker, consistently with Hirsch's prediction. Averaging the measurements, we get a difference of \SI{1.8}{Hz} at \SI{180}{pm} from point contact, corresponding to an average force gradient difference of \SI{0.13}{N/m}. However, this difference cannot be attributed to superconductivity, since the control experiment presented in Fig.~\ref{fig:df}(c) shows that a comparable effect is measured also on a non-superconductive Cu(110) sample.

From the latter data, we can estimate the overall accuracy of our measurements: the spectra at the two temperatures differ by \SI{2.2}{Hz} at \SI{180}{pm} from point-contact, which corresponds to a force gradient error $\delta \kts = \SI{0.17}{N/m}$. This value is the residual systematic error after having taken special care in order to characterize and account for possible instrumental effects, due in particular to the heating and cooling of the microscope. We considered specifically:

\paragraph*{Scanner calibration.} The position of the tip is controlled by a piezoelectric tube, whose calibration is the ratio between the tip apex displacement and the applied voltage, expressed in \SI{}{m/V}. This calibration depends on the temperature of the microscope, thus the $z$ measurements have been rescaled by measuring the height of a monoatomic step on the copper surface.\cite{Kroeger1977} The calibration changes by a factor of \SI{1.1359+-0.0035}{} going from \SI{4.4}{K} to \SI{9.5}{K}, and by a factor of \SI{1.1792+-0.0035}{} going from \SI{2.4}{K} to \SI{9.5}{K}. The stated precision corresponds to a relative standard uncertainty of \SI{3e-3}{}, and can be obtained by repeatedly measuring the step immediately before or after the $\Delta f (z)$ measurements, some tens of nanometers laterally away. If the calibration cannot be assessed close to the position of the spectroscopy, the slightly non-linear response of the piezo tube increases the uncertainty to about 1\%. This is the case for the measurements on the sputtered Nb surface shown in Fig.~\ref{fig:df}(b).

\paragraph*{Scanner creep and hysteresis.} The non-linearity of the piezo tube results in hysteresis in the tip displacement, and in creep -- the change of the tip position over time with an unchanged applied voltage. To minimize the effects of the former, we acquired the $\Delta f (z)$ measurements by sweeping $z$ always in the same direction, and we positioned the tip for a spectroscopy measurement by interpolating between images taken in the forward and in the backward scan directions. The creep decreases logarithmically over time after a tip displacement, so after having approached the tip we waited until there was no significant drift in the imaging over some minutes -- the timescale of the spectroscopy measurements. The accuracy of the tip-sample distance is key to our experiments, so before each spectroscopy the $z$-creep was measured by recording the voltage change needed to keep the tunneling current constant, and then compensated by subtracting the measured drift speed from the voltage controlling the $z$ position of the tip.

\paragraph*{qPlus calibration.} The amplitude calibration of the qPlus sensor does not change appreciably between \SI{4.4}{K} and \SI{9.4}{K}, as presented in Appendix \ref{app:a-calibration}.

\paragraph*{qPlus resonance frequency.} The resonance frequency of a qPlus sensor drifts with temperature.\cite{Pielmeier2014} We measured a change of \SI{-0.56}{Hz} going from \SI{4.4}{K} to \SI{9.5}{K}, and of \SI{-0.69}{Hz} from \SI{2.4}{K} to \SI{9.5}{K}. The frequency-shift data have been accordingly corrected.

\paragraph*{Bias voltage.} The frequency-shift measurements were acquired at a bias voltage giving zero tunneling current. In this way we avoid crosstalk effects between the AFM and the STM channels,\cite{Majzik2012} as well as changes in the electrostatic interaction between tip and sample due to thermoelectric voltages in the wires connecting them.

\paragraph*{Tip positioning.} 
In order to repeat the $\Delta f(z)$ spectra on the same point on the sample at the different temperatures, we used as a landmark an atomic-scale feature, such as an asperity on the sputtered niobium surface, or a defect on the copper surface. The $z$-axes of the measurements taken at different temperatures were aligned to a common point determined by a tunneling setpoint of $V = \SI{-20}{mV}$, $I = \SI{200}{pA}$. In order to have there the same tip-sample distance, this setpoint voltage was chosen well outside the superconductive gap, since the superconductive transition alters the electronic structure of the sample close to the Fermi level.

\subsection{Measurements at \SI{2.4}{K}}
\begin{figure}
	\includegraphics{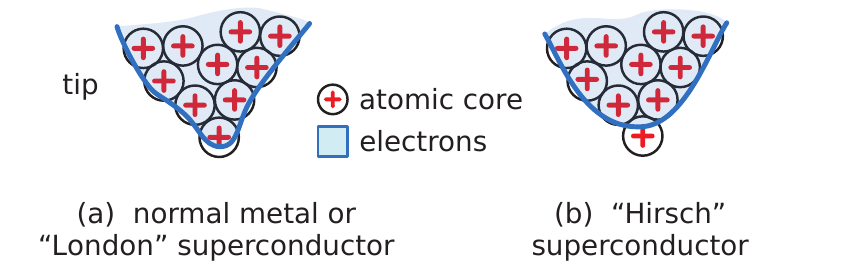}%
	\caption{\label{fig:tip} Schematic representation of a conductive tip apex and the origin of the tip dipole. The electron cloud does not follow the sharp curvature of the tip apex, giving rise to an electric dipole located at the apex -- the Smoluchowski effect. The increased screening length of the ``Hirsch'' superconductor results in a bigger dipole.}
\end{figure}

\begin{figure}
\includegraphics{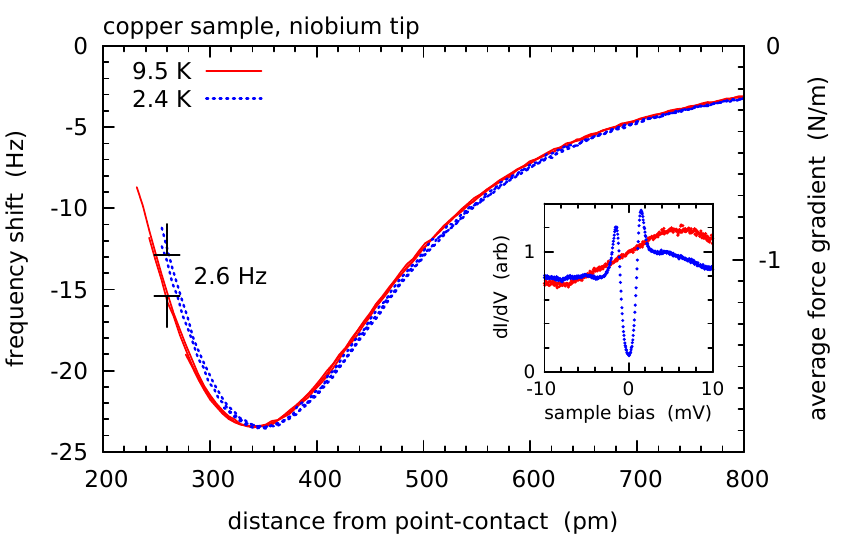}%
\caption{\label{fig:colder} $\Delta f(z)$ spectra at two different temperatures on Cu(111), acquired with a superconductive Nb tip. At $T = \SI{2.4}{K}$ the tip superconducts, as shown by the $\de I / \de V$ spectroscopy of the superconductive gap (inset). Similarly to Fig.~\ref{fig:df}, the $\Delta f(z)$ spectra at the two temperatures are different. These spectra are acquired on the same point on the surface, and multiple measurements are shown. The $\Delta f(z)$ measurements are acquired with an oscillation amplitude of \SI{100}{pm_{pk}}. The $\de I / \de V$ spectra are acquired at a tunneling setpoint $V = \SI{-20}{mV}, I = \SI{200}{pA}$ with a modulation voltage $\vm = \SI{200}{\micro V_{pk}}$ at $\fm = \SI{590}{Hz}$.}
\end{figure}

Effectively, Fig.~\ref{fig:df} shows that a systematic effect is present in our measurements, and that if the physics predicted by Hirsch are actually playing a role, this is smaller than the accuracy we are able to attain. Since the effect we are looking for is stronger the lower the temperature, we cooled our microscope to \SI{2.4}{K} by pumping on the helium bath and performed the experiment again, this time measuring $\Delta f(z)$ spectra on Cu(111) with a superconductive niobium tip. The tip was obtained by poking a tungsten tip into the annealed, oxygen-reconstructed niobium sample, as described in Ref.~\onlinecite{Ternes2006}. Measuring with a superconductive tip on a copper surface has two advantages: first, it is possible to precisely assess the calibration of the $z$-axis by measuring the height of a copper step directly after the $\Delta  f(z)$ measurement. Second, we observed that poking into niobium made the tip less reactive: in Fig.~\ref{fig:colder} the Pauli repulsion between the electron clouds of tip and sample is detectable at close tip-sample separations, due probably to an oxygen atom passivating the tip apex. Exchanging the role of tip and sample, however, changes also the physics we are interested in. Now the superconductive transition will increase the electrical screening length of the superconductive ``Hirsch'' tip, so its electron cloud will smooth out even more around the tip apex, giving rise to a bigger dipole -- see Fig.~\ref{fig:tip} -- and eventually to a \emph{stronger} tip-sample attraction, as opposed to the \emph{reduced} tip-sample attraction described in Figs.~\ref{fig:dipole} and \ref{fig:df}. Indeed, Fig.~\ref{fig:colder} shows that the spectra acquired at \SI{2.4}{K} are different from those acquired at \SI{9.5}{K}, at most by \SI{2.6}{Hz} at \SI{260}{pm} from point-contact. This corresponds to a force gradient difference of \SI{0.19}{N/m}, which is still comparable to what we observed in the control experiment of Fig.~\ref{fig:df}(c), and is thus not enough to confirm Hirsch's theory.

\section{Conclusions}
We attempted to test between two different theories describing the electrodynamics of superconductors: the traditional London theory and the theory proposed by J.~E. Hirsch. By means of AFM spectroscopy, we tried to detect a change in the electrostatic interaction between a metal tip and a surface when one of the two becomes superconductive. We observed a small effect, which is however below the accuracy of our measurements, and thus not enough to support Hirsch's hypothesis.  Since we are not able to provide a \emph{lower} estimate of the magnitude of the effect predicted by Hirsch, our measurements are not sufficient to disprove his hypothesis either. We can generally conclude that the superconductive transition does not affect the tip-sample interaction within our experimental accuracy of $\delta \kts = \SI{0.17}{N/m}$ at \SI{180}{pm} from point-contact.

Since the effect we are looking for increases dramatically for $T/T_c \lesssim 0.1$,\cite{Hirsch2012, Hirsch2015} further experiments should be conducted below \SI{1}{K}, or on a material with an higher $T_c$, such as \ce{Nb3Sn} or \ce{MgB2}. The accuracy of the measurements could also be greatly improved by using a magnetic field instead of the temperature to quench the superconductivity. Indeed, the main factor limiting our accuracy are the experimental difficulties associated with heating the microscope.

The applications of superconducting STM tips\cite{Pan1998} are not limited to the investigation of superconductor physics.\cite{Naaman2001, Giubileo2001, Rodrigo2004, Rodrigo2004a} Such tips have been used to increase the resolution of $\de I / \de V$
scanning tunneling spectroscopy (STS) by using the sharp edge of the superconducting gap to probe the electronic states
of the sample,\cite{Franke2011} to assess the instrumental resolution in STS,\cite{Rodrigo2004} and to measure local spin polarizations.\cite{Eltschka2014} We demonstrated here a superconductive qPlus sensor, which combines these possibilities with the measurement of forces at the nanoscale.

\begin{acknowledgments}
We thank the Deutsche Forschungsgemeinschaft for funding under the Sonderforschungsbereich 689. A.P. thanks J. Repp, J. E. Hirsch, C. Strunk, and M. Eschrig for useful discussion and suggestions, and F. Griesbeck and A. Merkel for assembling the filters used to improve the resolution of the $\de I/\de V$ measurements. We are grateful to A. J. Weymouth, L. L. Patera, F. Huber, J. Berwanger, S. Matencio, and D. Meuer for carefully proofreading the manuscript.
\end{acknowledgments}

\begin{figure*}
	\includegraphics{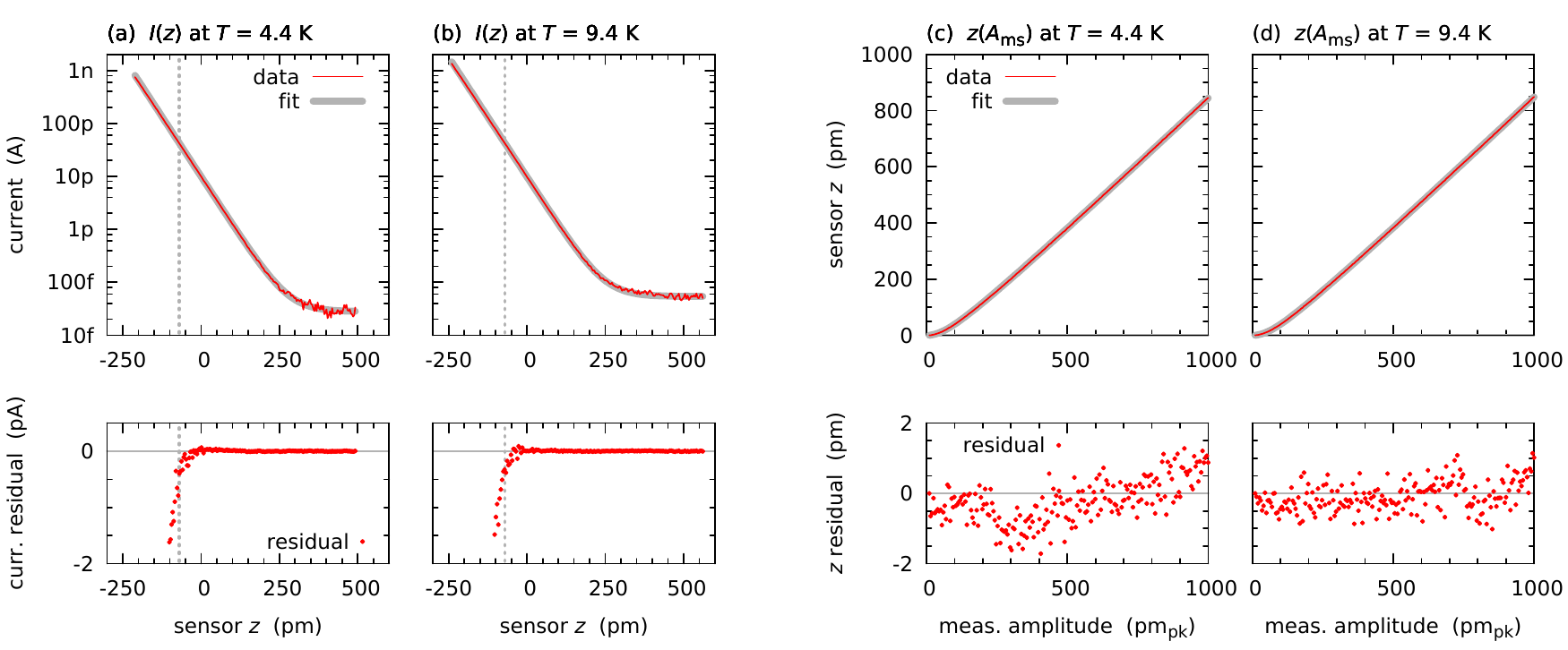}%
	\caption{\label{fig:iz-za} (a) and (b): $I(z)$ curves measured on Cu(110) at $T = \SI{4.4}{K}$ and $T = \SI{9.4}{K}$. Each curve is the average of 5 forward and 5 backward $z$ sweeps, taken with a measured oscillation amplitude $A_\mathrm{ms} = \SI{10}{pm_{pk}}$. $z = 0$ corresponds to a tunneling setpoint of $V = \SI{-20}{mV}$, $I = \SI{10}{pA}$. The function \eqref{eq:iz} has been fit to the data, including only the points with $z > \SI{-70}{pm}$ (dotted line), because the residuals shown in the lower panels indicate that close to the sample the $I(z)$ curves deviate from the exponential behavior. (c) and (d): $z(A_\mathrm{ms})$ curves measured on Cu(110) at  $T = \SI{4.4}{K}$ and $T = \SI{9.4}{K}$. Each curve is the average of one forward and one backward amplitude sweep, taken at a tunneling setpoint of $V = \SI{-20}{mV}$, $I = \SI{10}{pA}$, and starting from $A_{0 \, \mathrm{ms}} = \SI{10}{pm_{pk}}$. The function \eqref{eq:z3} has been fit to the data, the residuals are shown in the lower panels, and the fit results are reported in Tab.~\ref{tab:fit}.}
\end{figure*}

\appendix

\section{\label{app:a-calibration}Calibration of the qPlus sensor}

The amplitude calibration of the qPlus sensor is the ratio between the actual oscillation amplitude of the tip apex and the voltage output of the AFM amplifier, expressed in m/V. We describe here how we measured this calibration at  $T = \SI{4.4}{K}$ and at $T = \SI{9.4}{K}$, in order to check for a possible temperature dependence. The measurement consists in sweeping the oscillation amplitude $A$, while regulating the sensor position $z$ in order to keep the average tunneling current $I$ constant. The obtained $z(A)$ curve relates the oscillation amplitude, possibly miscalibrated, with the sensor position, which can be precisely calibrated by measuring a monoatomic step on the sample. For this reason, the sensor calibration can be determined by comparing these $z(A)$ curves with the expression derived in the following, with a precision limited by the precision of the scanner calibration.

The time-averaged tunneling current $I$ for a sensor at position $z$ oscillating with a peak amplitude $A$ between $z - A$ and $z + A$ is\cite{Berdunov2009}
\begin{equation} \label{eq:iza}
I(z, A) = I_0 \; e^{-2 \kappa z} \; \bes(2 \kappa A) ,
\end{equation}
where $\kappa$ is the decay constant of the tunneling current, $\bes$ is the modified Bessel function of the first kind of argument 0, and $I_0$ is the tunneling current at $z = 0$ and $A = 0$.  Rearranging, the sensor position can be expressed as
\begin{equation} \label{eq:z1}
z = \frac{1}{2 \kappa} \; \ln \left( \frac{ \bes(2 \kappa A) }{ I/I_0 } \right) .
\end{equation}
From Eq.~\eqref{eq:iza}, the ratio $I/I_0$ is
\begin{equation}
I/I_0 = e^{-2 \kappa z} \; \mathcal{I}_0(2 \kappa A) ,
\end{equation}
and since the zero of the $z$ position is arbitrary, we can define $z = 0$ at the beginning of the amplitude sweep, where the amplitude is $A_0$. In this way the previous equation simplifies to
\begin{equation}
I/I_0 = \bes(2 \kappa A_0) ,
\end{equation}
and equation \eqref{eq:z1} becomes
\begin{equation} \label{eq:z2}
z = \frac{1}{2 \kappa} \; \ln \left( \frac{ \bes(2 \kappa A) }{ \bes(2 \kappa A_0) } \right) .
\end{equation}
If the amplitude calibration is wrong, the measured amplitude $A_\mathrm{ms}$ is related to the true amplitude $A$ by the linear relation
\begin{equation}
A = \eta A_\mathrm{ms} ,
\end{equation}
which defines the dimensionless miscalibration factor $\eta$. In terms of $A_\mathrm{ms}$, Eq.~\eqref{eq:z2} becomes
\begin{equation} \label{eq:z3}
z = \frac{1}{2 \kappa} \; \ln \left( \frac{ \bes(2 \kappa \eta A_\mathrm{ms}) }{ \bes(2 \kappa \eta A_{0 \, \mathrm{ms}}) } \right) .
\end{equation}

Equation~\eqref{eq:z3} can be fit to the measured $z(A_\mathrm{ms})$ curves by varying only $\eta$, since the initial amplitude $A_{0 \, \mathrm{ms}}$ is obtained from the experimental data, and the decay constant $\kappa$ has been determined from an independent $I(z)$ measurement described in the following. The fits employed the trust-region reflective algorithm implemented in the Curve Fitting Toolbox of \textsc{matlab} 2014b, and the results are presented in Fig.~\ref{fig:iz-za} and Tab.~\ref{tab:fit}. In particular, $\eta(\SI{9.4}{K})$ does not significantly differ from $\eta(\SI{4.4}{K})$.

\begin{table}
	\caption{\label{tab:fit}%
		Fit results for the two temperatures. The reported standard uncertainties combine the uncertainty from the fit procedure with the uncertainty of the scanner calibration.}
	\begin{ruledtabular}
		\begin{tabular}{cccc}
			\multirow{2}{*}{$T$ (\si{K})}& \multicolumn{1}{c}{from $z(A_\mathrm{ms})$} & \multicolumn{2}{c}{from $I(z)$} \\
			\cline{2-2} \cline {3-4}
			&
			$\eta$ &
			$\kappa$ (\si{1/nm})&
			$I_0$ (\si{pA})\\
			\colrule
			4.4 & 0.9608(20) & 10.425(23) & 9.771(14) \\
			9.4 & 0.9646(35) & 10.404(42) & 9.613(32) \\
		\end{tabular}
	\end{ruledtabular}
\end{table} 

\paragraph*{Decay constant of the tunneling current.} Taking into account the offset $I_\mathrm{ofs}$ of the current amplifier, equation \eqref{eq:iza} becomes
\begin{equation} \label{eq:iz}
I(z) = I_0 \; \bes(2 \kappa A) \; e^{-2 \kappa z} + I_\mathrm{ofs} .
\end{equation}
This function was fit to the $I(z)$ curves measured at \SI{4.4}{K} and at \SI{9.4}{K} by varying $I_0$ and $\kappa$, whereas $I_\mathrm{ofs}$ and $A$ were measured directly. As shown in Fig.~\ref{fig:iz-za} and Tab.~\ref{tab:fit}, also the decay constant $\kappa$ does not change significantly between the two temperatures.  Since these fits depend on the oscillation amplitude $A$, they had to be repeated after having determined the amplitude miscalibration factor $\eta$ in order to get the correct value of $I_0$, whereas at our precision $\kappa$ is independent form the exact value of $A$.

\inputencoding{latin2}
\bibliography{sc-paper}
\inputencoding{utf8}

\end{document}